\documentstyle[12pt]{article}
\newcommand{\be}{\begin{equation}}
\newcommand{\ee}{\end{equation}}
\begin{document}

\title{Simple Linearizations of the Simple Pendulum for Any Amplitude\footnote{mmolina@abello.dic.uchile.cl}}
\author{{\bf M. I.  Molina}
\vspace{1 cm}
\and
\and
Facultad de Ciencias, Departamento de F\'{\i}sica, Universidad de Chile\\
Casilla 653, Las Palmeras 3425, Santiago, Chile.}
\date{}
\maketitle
\baselineskip 24 pt
\vspace{1.0in}
\begin{center}
{\bf Abstract}
\end{center}

\noindent
Two simple, interpolatory-like linearizations are shown for the simple
pendulum which can be used for any initial amplitude.
\vspace{1cm}

\noindent
\centerline{\em Key words:\ \ pendulum, linearization.}\\
\centerline{\em PACS:\ \ 07.10.Y, 02.60.E}

\newpage
%%%%%%%%%%%%%%%%%% MAIN BODY %%%%%%%%%%%%%%%%%%%%%%%%%%%%%%%%%%%%%%%%%

A commonplace approximation when teaching the motion of a simple pendulum
is to assume a small initial amplitude, so that the equation of motion can
be safely
approximated with the one corresponding to a harmonic oscillator,
leading to an amplitude-free period. If one wishes to consider the case
of not necessarily small initial amplitudes one is obliged to use the
full formula for the period of the pendulum, involving an elliptic integral 
or its perturbation series$^{1}$ or numerical methods. I describe below, a simple way around this
difficulty that allows one to obtain an amplitude-dependent period for the
pendulum that can be used for any initial amplitude.

Let us consider a simple pendulum of length $l$ in a (uniform)
gravitational field $g$. The equation of motion is
\be
{d^{2} \Theta\over{dt^{2}}} = -\left({g\over{l}}\right) \sin(\Theta) \label{eq:1}
\ee
If the initial amplitude is small, we can approximate $\sin(\Theta)$
by $\Theta$ (linearization), which leads to the approximate equation
$(d^{2} \Theta/dt^{2}) = -(g/l) \Theta$, i.e., the harmonic oscillator
equation with period $T_{0} = 2 \pi \sqrt{l/g}$.
What we want to do is to extend this linearization procedure in such a way
as to be able to use it even at {\em large initial amplitudes}. To do that,
we have to replace the $\sin(\Theta)$ term by something of the form 
$F(\Theta_{0}) \Theta$, where $F(\Theta_{0})$ is an amplitude-dependent
coefficient. This term will compensate for the fact that the force is
not really linear in $\Theta$. Thus, for small initial amplitudes where the
usual linearization is valid, $F(\Theta_{0}) \approx 1$, while for amplitudes
near $\pi$, where the force on the pendulum is small, $F(\Theta_{0})\approx
0$. In addition, $F(\Theta_{0})$ must be even in $\Theta_{0}$ since by
symmetry, the period of the pendulum can only depend on the magnitude of the
initial amplitude, $|\Theta_{0}|$. A simple $F(\theta_{0})$ that obeys these requirements is
\be
F(\Theta_{0}) = \left({\sin(\Theta_{0})\over{\Theta_{0}}}\right)^{\alpha} \label{eq:0}
\ee
where $\alpha$ is a number to be chosen judiciously (see below). With
the above, the equation of motion is
\be 
{d^{2} \Theta\over{dt^{2}}} + \left( {g\over{l}} \right)
\left( {\sin(\Theta_{0})\over{\Theta_{0}}} \right)^{\alpha} \Theta =0
\ee
and leads to a period $T = T_{0} (\Theta_{0}/\sin(\Theta_{0}))^{\alpha/2}$,
The parameter $\alpha$ can be chosen in a variety of ways. We can,
for instance, impose that the main corrective term in the low-amplitude
expansion of T, coincide with the leading corrective term for the
{\em exact} period$^{1}$:
\be
T_{exact} = (2/\pi)\ T_{0}\ \int_{0}^{\pi/2} {d\phi\over{(1-k^{2}
\sin(\phi)^{2})^{1/2}}}
\label{eq:exact}
\ee
with $k = \sin(\Theta_{0}/2)$. For small $\Theta_{0}$, we have
\be
T_{exact} = T_{0} ( 1 + (1/16) \Theta_{0}^{2} + ...) \label{eq:2}
\ee
On the other hand, in our scheme we have
\be
T = T_{0} ( 1 + (\alpha/12) \Theta_{0}^{2} + ...)  \label{eq:3}
\ee
Comparison of Eqs.(\ref{eq:2}) and (\ref{eq:3}) determines
$\alpha = 3/4$, and $T$ becomes
\be
T = T_{0} \left( {\Theta_{0}\over{\sin(\Theta_{0})}} \right)^{3/8}
\label{eq:final}
\ee

In the same spirit, the reader can check that the choice
\be
F(\Theta_{0}) = \left[ 1 - \left({\Theta_{0}\over{\pi}}\right)^{2} \right]^{\pi^{2}/8}
\ee
also satisfy our requirements and leads to a period
\be
T = T_{0} \left[ 1- \left( {\Theta_{0}\over{\pi}} \right)^{2} \right]^{- \pi^{2}/16}
\label{eq:4}
\ee
which, in the low
amplitude limit, coincides with Eq.(\ref{eq:2}), although its fit over
all the angular range is not as good as with the first choice,
Eq.(\ref{eq:0}).

Figure 1 shows a comparison between the periods for the exact case,
Eq.(\ref{eq:exact}), the main perturbative correction to the low-amplitude
case, Eq.(\ref{eq:2}) and our two interpolatory approximations,
Eqs.(\ref{eq:final}) and (\ref{eq:4}). The agreement using the
interpolation scheme is quite satisfactory, remaining at
$\Theta=2\ (114.6^{0})$
within $1\%$ and $4\%$ from the exact values for interpolations (\ref{eq:final}) and (\ref{eq:4}),
respectively.

\vspace{4cm}
         
%%%%%%%%%%%%%%%%%%%%%%%%%%%%%%%%%%%%%%%%%%%%%%%%%%%%%%%%%%%%%%%%%%%%%%
\centerline{\bf References}
\vspace{1cm}

\noindent
$^{1}$ J. B. Marion and S. T. Thornton, {\em Classical Dynamics of Particles
and Systems}, third edition (Saunders, 1988), pp. 143-147.
\newpage

\centerline{{\bf Captions List}}

\vspace{2cm}

\noindent {\bf Fig.1 :}\ \ Comparison of periods for the simple pendulum
obtained from the exact solution, perturbative, interpolation 
scheme (\ref{eq:final}) and interpolation scheme (\ref{eq:4}).

\newpage

%  Figures follow

%%%%%%%%%%%%% 	FIG. 1   %%%%%%%%%%%%%%%%%%%%%%%%%%%%%%%%%%%%%
\begin{figure}[p]
\begin{center}
\leavevmode
\hbox{
\includegraphics{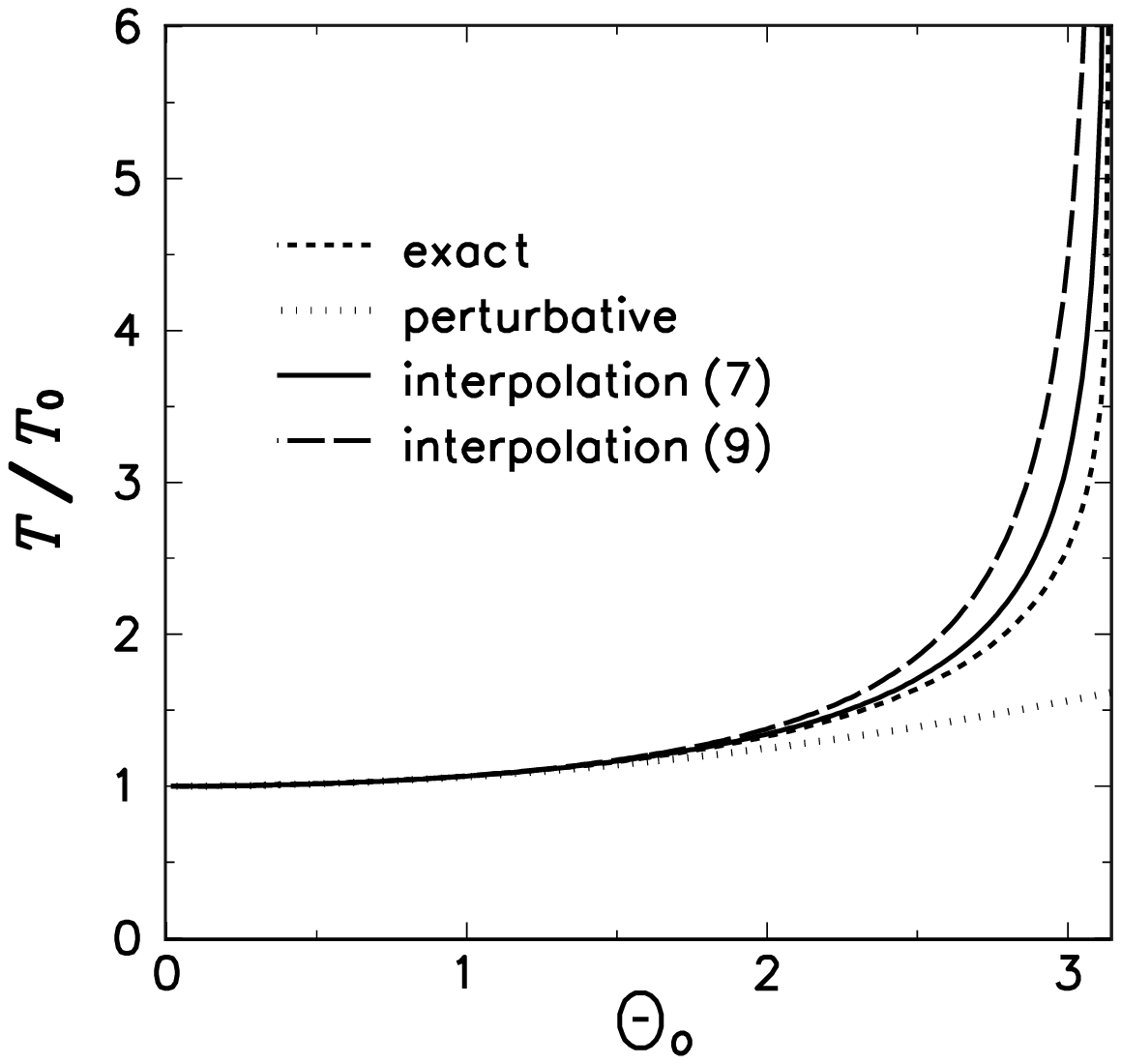}}
\end{center}
\vspace{2.3 in}
\label{figure1}
\vspace{10.0cm}
%\centerline{Fig.1}
\end{figure}

\end{document}